\documentclass[conference]{IEEEtran}
\IEEEoverridecommandlockouts
\usepackage{cite}
\usepackage{amsmath,amssymb,amsfonts}
\usepackage{algorithmic}
\usepackage{graphicx}
\usepackage{textcomp}
\usepackage{xcolor}
\def\BibTeX{{\rm B\kern-.05em{\sc i\kern-.025em b}\kern-.08em
    T\kern-.1667em\lower.7ex\hbox{E}\kern-.125emX}}
\begin{document}

\title{Implementation and Applications of WakeWords Integrated with Speaker Recognition: A Case Study
}
\author{
\IEEEauthorblockN{
    \begin{minipage}[t]{0.45\textwidth}
    \centering
    Alexandre Costa Ferro Filho\\
    \textit{Computer Institute} \\
    \textit{Goias Federal University}\\
    Goiânia, Brazil \\
    alexandre\_ferro@discente.ufg.br
    \end{minipage}
    \hfill
    \begin{minipage}[t]{0.45\textwidth}
    \centering
    Elisa Ayumi Masasi de Oliveira\\
    \textit{Computer Institute} \\
    \textit{Goias Federal University}\\
    Goiânia, Brazil \\
    ayumi@discente.ufg.br
    \end{minipage}
    \\
    \\
    \vspace{10mm} 
    \begin{minipage}[t]{0.45\textwidth}
    \centering
    Iago Alves Brito\\
    \textit{Computer Institute} \\
    \textit{Goias Federal University}\\
    Goiânia, Brazil \\
    iagoalves@discente.ufg.br
    \end{minipage}
    \hfill
    \begin{minipage}[t]{0.45\textwidth}
    \centering
    Pedro Martins Bittencourt\\
    \textit{Computer Institute} \\
    \textit{Goias Federal University}\\
    Goiânia, Brazil \\
    bittencourtpedro@discente.ufg.br
    \end{minipage}
}
}
\maketitle

\begin{abstract}

This paper explores the application of artificial intelligence techniques in audio and voice processing, focusing on the integration of wake words and speaker recognition for secure access in embedded systems. With the growing prevalence of voice-activated devices such as Amazon Alexa, ensuring secure and user-specific interactions has become paramount. Our study aims to enhance the security framework of these systems by leveraging wake words for initial activation and speaker recognition to validate user permissions. By incorporating these AI-driven methodologies, we propose a robust solution that restricts system usage to authorized individuals, thereby mitigating unauthorized access risks. This research delves into the algorithms and technologies underpinning wake word detection and speaker recognition, evaluates their effectiveness in real-world applications, and discusses the potential for their implementation in various embedded systems, emphasizing security and user convenience. The findings underscore the feasibility and advantages of employing these AI techniques to create secure, user-friendly voice-activated systems.
\end{abstract}

\begin{IEEEkeywords}
Audio Processing, Wake Words, Speaker Recognition, Embedded Systems, Synthetic Data

\end{IEEEkeywords}

\section{Introduction}

In recent years, the popularization of voice-activated devices has transformed the way users interact with technology. Devices like Amazon Alexa, Google Home, and Apple Siri have popularized voice as a convenient interface for accessing information, controlling smart home devices, and performing various tasks. However, as the adoption of these devices grows, the concern for security and privacy becomes an important discussed topic, since the unauthorized access to these systems can lead to breaches of personal information and unauthorized control of connected devices \cite{lau2018alexa}.

To address these concerns, the integration of wake words and speaker recognition in voice-activated systems has emerged as a promising solution. Wake words, such as "Alexa" or "Hey Siri" serve as verbal triggers that activate the device and prepare it to receive commands. Speaker recognition, on the other hand, involves identifying and verifying the identity of the speaker based on their unique vocal characteristics. Together, these technologies can enhance the security framework of voice-activated systems by ensuring that only authorized users can interact with the device.

To develop the wake word detection model, synthetic data was employed to augment the training dataset. The use of synthetic data is a crucial aspect of our approach, as it allows for the creation of diverse and extensive training examples, which are essential for building robust and accurate models. This technique is particularly valuable when dealing with the scarcity of real-world data, especially for specific wake words or in scenarios where data privacy concerns limit the availability of authentic recordings. The synthetic data generation process involved creating varied and realistic audio samples that mimic the characteristics of natural speech, thus providing the model with a broad range of vocal patterns and environmental conditions to learn from.

In the following sections, we provide an overview of the current state of wake word detection and speaker recognition technologies, describe the methodology and experimental setup used in our research, present the results of our experiments, and discuss the implications of our findings. Finally, we conclude with a discussion of future research directions and the potential for further advancements in this field.

\section{Background}

\subsection{Voice-Activated Systems and Security Concerns}
Voice-activated systems have become increasingly prevalent in everyday life, offering hands-free convenience and intuitive interfaces for interacting with technology \cite{smarthome}. These systems rely on Natural Language Processing (NLP) and machine learning (ML) to understand and respond to user commands. However, the widespread use of voice commands also introduces significant security and privacy challenges. Unauthorized users can potentially access sensitive information or control connected devices, leading to security breaches.

\subsection{Wake Words}

Wake words are predefined phrases used to activate voice-activated systems, signaling that the device is ready to receive commands \cite{VoiceActivatedSystems}. This mechanism serves as an initial security layer by requiring a specific trigger before any interaction can occur. The detection of wake words must be both precise and efficient, ensuring that the system responds only to intentional activations while minimizing false positives and negatives.

\subsection{Speaker Recognition}

Speaker recognition technology identifies and verifies individuals based on their unique vocal characteristics. This process involves two main stages: speaker identification and speaker verification. Speaker identification determines the identity of the speaker from a group of known individuals, while speaker verification confirms whether the speaker matches a claimed identity \cite{speechrecognizer}.

Speaker recognition enhances security in voice-activated systems by restricting access to authorized users. This capability is critical for applications involving sensitive information or critical functions, where it is essential to ensure that only specific individuals can issue commands.

\section{Methodology}

\subsection{Data Collection and Synthetic Data Generation}
In developing robust wake word detection and speaker recognition systems, data quality and diversity are crucial. For wake word detection, the training dataset included both real and synthetic audio samples. The real data comprised a small variety of wake word samples from 3 different speakers. However, synthetic data was generated to supplement the real dataset, representing the majority part of the dataset. This synthetic data was created using text-to-speech (TTS) systems and audio augmentation techniques, such as described in \cite{ko2015data}, producing a wide range of samples that mimic natural variations in pronunciation, background noise, and recording conditions. This approach is particularly useful in our scenario where authentic data is limited. The inclusion of synthetic data allows for more comprehensive training, improving the model's ability to generalize across different speakers and environmental conditions.

\subsection{Wake Word Detection Model}
The wake word detection system was built using a deep learning architecture designed to handle the variability and complexity of speech data. The model architecture incorporated a Convolutional Neural Network (CNN) for feature extraction. The use of CNN layers enabled the model to learn hierarchical representations of the audio data, which is crucial for accurately detecting the presence of wake words, where, in this case, the wake word "Hey, Gris!" was chosen.

To enhance the performance of wake word models, we investigated and found that including the use of synthetic data is beneficial to model performance when using the Google proposed model 
c. 

\subsection{Speaker Recognition System}
\begin{figure}[h!]
    \centering
    \includegraphics[width=0.4\textwidth]{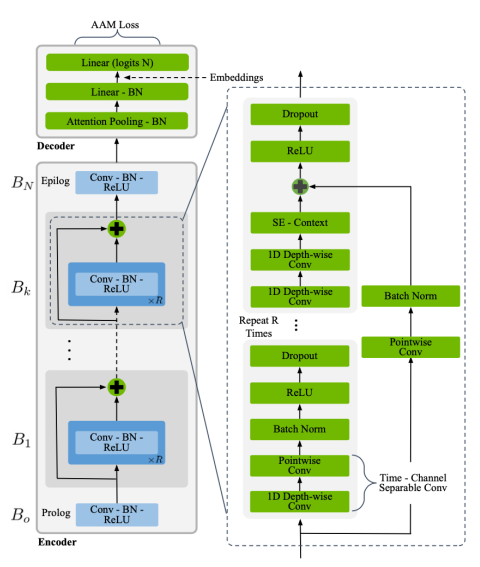} 
    \caption{NVIDIA TitaNet model architecture.}
    \label{fig:titanet}
\end{figure}

The speaker recognition component aimed to authenticate users by verifying their identity based on voice characteristics. For this purpose, we employed the Titanet architecture \cite{titanet}, a state-of-the-art model designed for robust speaker recognition. Titanet, known for its efficiency and accuracy, utilizes an end-to-end deep learning approach that integrates feature extraction and classification within a unified framework. This architecture, represented in figure \ref{fig:titanet}, is particularly well-suited for processing variable-length audio segments and capturing the nuanced vocal characteristics that distinguish different speakers.

Titanet consists of multiple layers, including residual blocks and attention mechanisms, which enhance the model's ability to focus on relevant features in the audio data. The training process for Titanet involved a large dataset of labeled speaker data, ensuring that the model could learn to accurately differentiate between speakers. The use of attention mechanisms allowed Titanet to dynamically weigh different parts of the input audio, improving its capacity to handle variations in speaking style and environment.

To enhance security and reduce the likelihood of spoofing or impersonation, the system incorporated anti-spoofing measures, including the detection of synthetic voices and other anomalies that could indicate unauthorized access attempts.

\section{Results}

\subsection{Empirical Evaluation}

The effectiveness of our systems was assessed through an empirical evaluation conducted in real-world environments. The primary focus was on evaluating the response time and reliability of both the wake word detection and speaker recognition models under various conditions, including different noise levels and speaker variations. This practical testing aimed to simulate real-world scenarios where these systems would typically be deployed, such as in homes or offices.

Preliminary results indicate that the system effectively recognizes wake words and accurately identifies speakers, even in challenging acoustic environments. The use of synthetic data during training has significantly enhanced the model's robustness, allowing it to generalize well across different voices and background noises. However, while these results are promising, they were primarily based on qualitative observations. Future evaluations would benefit from a more detailed quantitative analysis, which could offer a clearer understanding of the system's performance and highlight specific areas for improvement. This approach would provide a more comprehensive assessment of the system's capabilities and help guide further refinement and optimization efforts.

\section{Conclusion}

This study has demonstrated the potential of integrating artificial intelligence techniques, such as wake word detection and speaker recognition, to enhance the security of voice-activated systems in embedded environments. By leveraging wake words for system activation and speaker recognition for user verification, the approach restricts system usage to authorized individuals, thereby mitigating the risks of unauthorized access. The use of synthetic data in training the wake word detection model was crucial, enabling the system to generalize across diverse speakers and environmental conditions, while the Titanet architecture provided robust speaker verification.

The empirical evaluation confirmed the effectiveness of these AI-driven methodologies in real-world settings, showcasing their ability to create secure and user-friendly voice-activated systems. The integration with ROS and Docker streamlined the development and deployment processes, ensuring adaptability across various hardware platforms. Future research could focus on optimizing models for low-resource environments, enhancing anti-spoofing measures, and expanding the use of synthetic data to further secure and refine voice-activated technologies.

\bibliographystyle{plain} 
\bibliography{references} 

\end{document}